\documentstyle[
	aps, 
	epsf,
	twocolumn, 
	prb, 
	floats 
	]{revtex}

\begin{document}

\twocolumn[\hsize\textwidth\columnwidth\hsize\csname @twocolumnfalse\endcsname 

\title{
	Epitaxial growth of Cu on Cu(001): 
	 experiments and simulations
}
\smallskip
\author{Itay Furman and Ofer Biham}
\address{
Racah Institute of Physics,
The Hebrew University,
Jerusalem 91904,
Israel}
\smallskip
\author{Jiang-Kai Zuo\footnote{
present address:
Semiconductor Product Sector,
Motorola, Inc., MD-M350,
Mesa, AZ 85202, USA
},
Anna K. Swan\footnote{
present address:
Department of Electrical and Computer Engineering,
Boston University,
Boston, MA  02215, USA
}
and John F. Wendelken}
\address{
Solid State Division,
Oak Ridge National Laboratory,
Oak Ridge, TN 37831,
USA}
\date{\today}
\maketitle
\bigskip
\begin{center}
(accepted for publication in {\it Phys.\ Rev.\ B, Rapid Communications})
\end{center}
\smallskip

\begin{abstract}
A quantitative comparison between experimental and
Monte Carlo simulation results
for the epitaxial growth of Cu/Cu(001) in the submonolayer regime
is presented.
The simulations take into account a complete set of hopping processes whose
activation  energies are derived from
semi-empirical calculations using the embedded-atom method.
The island separation is measured as a function of the incoming flux
and the temperature.
A good {\it quantitative} agreement between the experiment and simulation
is found for the island separation,
the activation energies for the dominant processes,
and the exponents that characterize the growth.
The simulation results are then analyzed at lower coverages, which are
not accessible experimentally, providing good agreement with 
theoretical predictions as well.
\end{abstract}

\smallskip
\smallskip

] 
\narrowtext 

\smallskip


The growth  of a newly deposited layer (adlayer)
of a thin metal film in molecular-beam epitaxy (MBE)
through island nucleation  involves three kinetic regimes
with respect to the coverage $\theta$:
(a) {\it Nucleation regime},
dominated by island nucleation;
(b) {\it Aggregation regime},
where newly deposited atoms (adatoms) are mostly captured by existing islands.
In this regime the system is in a quasi steady-state,
a property that is employed in the scaling theory;
(c) {\it Coalescence regime},
where islands are large enough to interact and merge.

Experiments using scanning tunneling microscopy (STM),%
\cite{ref:STM4T,ref:STM4FT,Stroscio1994}
low-energy electron diffraction%
\cite{Zuo1994S,Durr1995,Swan1997S,Bardotti1998S}
and other methods,%
\cite{Ernst1992}
made in attempt to study the aggregation regime,
reveal an exponential dependence
of the island density $N$, on the inverse temperature $1/T$,%
\cite{ref:STM4T,ref:STM4FT,Durr1995,Bardotti1998S,Ernst1992}
as well as a power law dependence on the flux $F$.%
\cite{ref:STM4FT,Stroscio1994,Zuo1994S,Swan1997S,Ernst1992}
This behavior can be described by the following formula:%
\cite{ref:nucleation}
\begin{equation}
	\label{eq:nucleation}
	\ell^{-2} \sim \left( F \over \nu \right) ^\gamma
		\exp \left( E_{\rm eff} / k_{\rm B} T \right),
\end{equation}
where $\ell$ is the mean island separation related to $N$ through
$N \sim \ell^{-2}$.
In the experimental-data analysis,
$\gamma$, $\nu$ and $E_{\rm eff}$ are fitting parameters,
where $E_{\rm eff}$ is a measure of
the activation energy for the rate-limiting move,
$\nu$ is the attempt rate for that move
and $\gamma$ is a scaling exponent.
The theoretical challenge is, therefore,
to relate $\gamma$, $\nu$ and $E_{\rm eff}$
to the fundamental processes at the atomic level.

\begin{figure}
\vspace{-50mm}
\centerline{
\hskip -30mm
\vbox{\epsfxsize=210mm \epsfbox {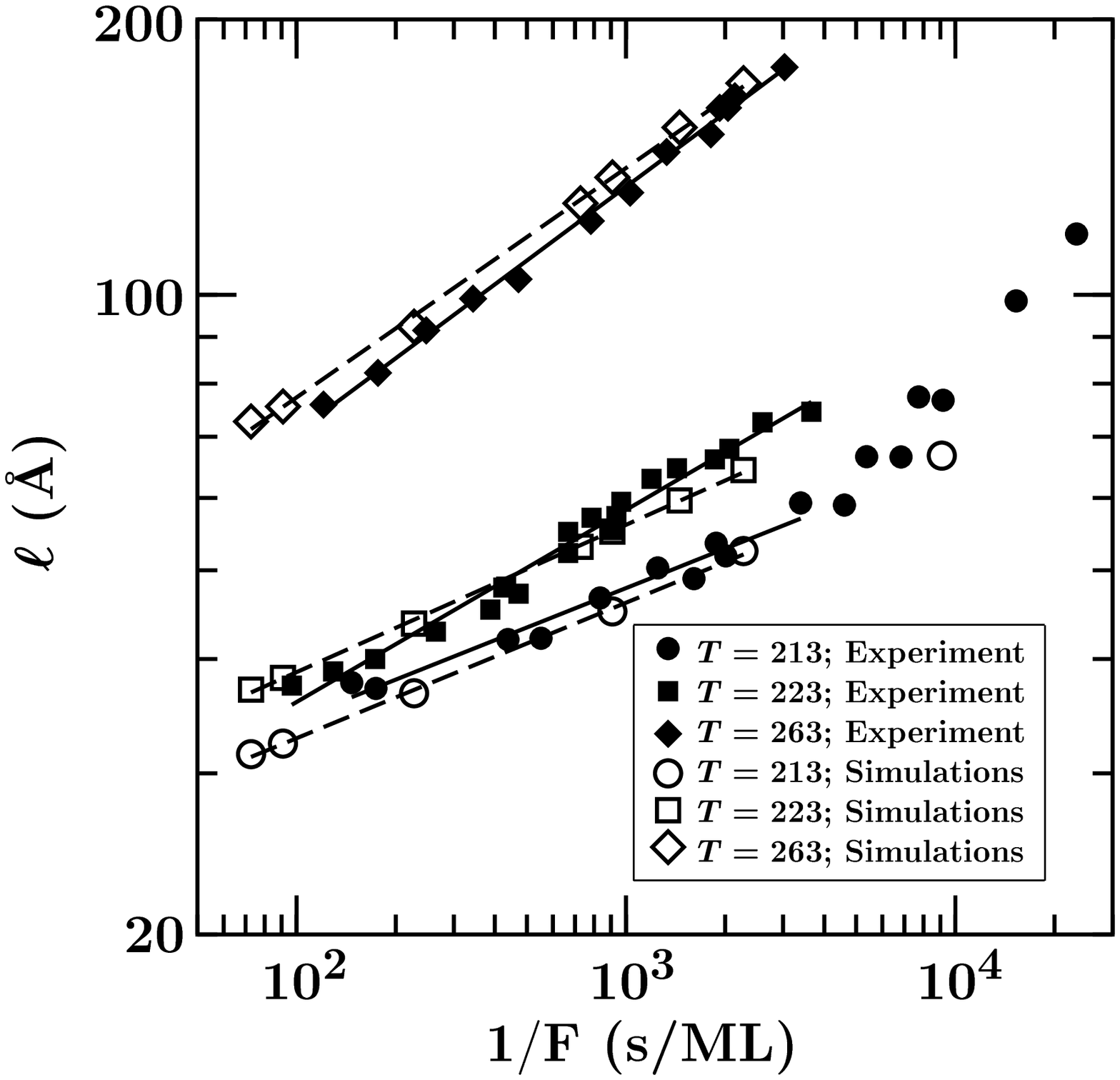}}
}
\vspace{-80mm}
\caption{
Comparison of experimental results (full symbols)
and simulation results (empty symbols)
for the island separation $\ell$ vs.\ the inverse flux $1/F$,
at three temperatures $T=213$, $223$ and $263$~K.
The coverage is $\theta = 0.3$~ML.
The solid lines represent fits to
Eq.~(\protect\ref{eq:nucleation}).
The resultant values of the exponent $\gamma$ are presented in
Table~\protect\ref{tab:gamma}.
}
\label{fig:gamma}
\end{figure}

\begin{figure}
\vspace{-50mm}
\centerline{
\hskip -30mm
\vbox{\epsfxsize=210mm \epsfbox {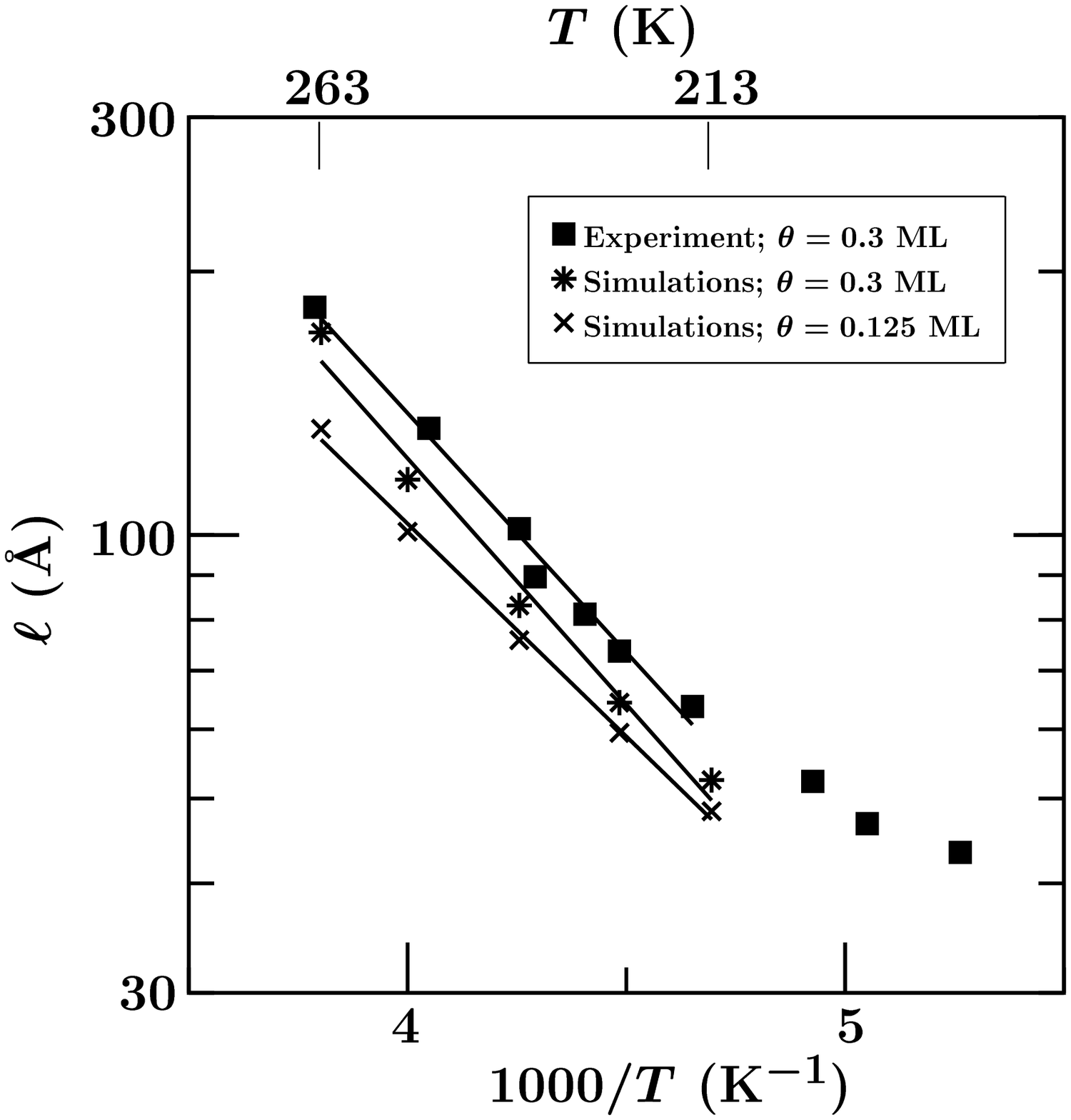}}
}
\vspace{-80mm}
\caption{
The island separation  $\ell$ vs.\ $1/T$,
obtained from the experiment for $F=3.21 \times 10^{-4}$~ML$\cdot$s$^{-1}$
and simulation for $F=4.8 \times 10^{-4}$~ML$\cdot$s$^{-1}$
at coverage $\theta = 0.3$~ML.
We also present the simulation results of the same runs
but at an earlier stage when $\theta = 0.125$~ML.
The solid lines represent fits to 
$\ell \sim \exp \left( - E_{\rm eff} / 2 K_{\rm B} T \right)$,
with $E_{\rm eff}= 0.108 \pm 0.005$, $0.112 \pm 0.008$
and $0.096 \pm 0.003$~eV for the
experiment, and the simulations at $\theta=0.3$ and $0.125$~ML,
respectively.
}
\label{fig:L2invT}
\end{figure}

In this paper we show that kinetic Monte Carlo (MC) simulations
based on energy barriers obtained via the embedded-atom method (EAM),
provide a good quantitative description of
diffusion and growth on Cu(001).
To this end we performed kinetic MC simulations of island growth
under conditions identical to those employed experimentally
in Refs.~\onlinecite{Zuo1994S,Durr1995,Swan1997S}.
In these experiments
spot-profile analysis of low-energy electron diffraction (SPA-LEED)
was used to measure the mean separation 
$\ell$
between Cu islands on Cu(001).
The measurements were taken 
as a function of the flux $F$
for three temperatures, $T = 213$, $223$ and $263$ K
(Fig.~\ref{fig:gamma}),
and as a function of the temperature $T$
at a constant flux $F = 3.21 \times 10^{-4}$~ML $\!\cdot\!$ s$^{-1}$
(Fig.~\ref{fig:L2invT}).

In our kinetic MC simulations atoms are randomly deposited
on a square-lattice substrate of $250 \times 250$ sites
that corresponds to a terrace width of $\sim 64$ nm.%
\cite{note:terrace_width}
These atoms attach irreversibly to the surface
and hop as random walkers to unoccupied nearest-neighbor (NN) sites.
Each hop involves an
activation energy-barrier, $E_n$, that
depends on the configuration of occupied and unoccupied adjacent sites
in the $3 \times 3$ square around the hopping atom.%
\cite{Voter1986,Furman1997,Biham1998S}
When an atom is deposited on top of an island
it is incorporated at a random position along the island perimeter.
The nucleation of a second layer is thus suppressed,
which is a good approximation for small island sizes at low coverages.
The hopping-rate (in units of hops per second), $h_n$,
for some configuration $n$ is%
\cite{note:barriers}
\begin{equation}
\label{eq:hoppingrate}
	h_n = \nu \cdot \exp(-E_n / K_{\rm B} T)
	,
\end{equation}
where $\nu$ is the attempt rate common to all moves
and $E_n$ is calculated using the EAM.%
\cite{Daw1983,Mehl1999,Biham1998S}

We have reconstructed the experimental conditions in our
kinetic MC simulations
and obtained for each experimental curve in
Figs.~\ref{fig:gamma} 
and 
\ref{fig:L2invT}
a corresponding simulated curve.
Each simulated data point is an average over 20 runs.
In 
Fig.~\ref{fig:gamma} 
we present the experimental results
and the corresponding simulation results,
for the island separation $\ell$ vs.\ $1/F$
at $T = 213$, $223$ and $263$~K.
All data points were taken at $\theta = 0.3$~ML.%
\cite{note:coverage}
Clearly, there is a good agreement between the experiment
and simulation.
The curves follow a  
power-law behavior according to
Eq.~(\ref{eq:nucleation})
in a range of one and a half decades,
from which the exponent $\gamma$ is extracted.
\cite{note:Sshape}
The results for the exponent $\gamma$ are presented in 
Table~\ref{tab:gamma}.%
\cite{note:range4fits}
It is found to increase in the range
$\gamma = 0.3$ -- $0.5$ as the temperature is raised,
suggesting that the system undergoes a crossver in its kinetic behavior.%
\cite{ref:crossover}
Based on scaling theory, one may conclude from the results
$\gamma \simeq 1/3$
that {\em dimers} and larger islands are stable and {\em immobile},
and adatoms are the {\em only} mobile entities on the surface.%
\cite{ref:nucleation}
However, care should be exercised when drawing such conclusions.
Specifically, in our case the value
$\gamma \simeq 1/3$
is found {\em in spite of significant dimer mobility}
as we show later.
\begin{table}
\vskip-\lastskip
\begin{tabular}{cr@{}l@{${}\pm{}$}r@{}l
			r@{}l@{${}\pm{}$}r@{}l
				r@{}l@{${}\pm{}$}r@{}l}
	& \multicolumn{12}{c}{$\gamma$}
	\\ \cline{2-13}
	$T$ (K) &
		\multicolumn{4}{c}{experiment} &
		\multicolumn{4}{c}{simulations} &
		\multicolumn{4}{c}{simulations} 
	\\
	&	\multicolumn{4}{c}{$\theta = 0.3$ (ML)} &
		\multicolumn{4}{c}{$\theta = 0.3$ (ML)} &
		\multicolumn{4}{c}{$\theta = 0.125$ (ML)} 
	\\	\hline

	213 &  & 0.28 & 0.02 &  &  & 0.30 & 0.01  &  &  & 0.309  & 0.006  &
	\\
	223 &  & 0.42 & 0.02 &  &  & 0.326  & 0.006  &  &  & 0.319  & 0.006  &
	\\
	263 &  & 0.54 & 0.01 &  &  & 0.51  & 0.01  &  &  & 0.396  & 0.006  &
	\\
\end{tabular}
\caption{
The values for the exponent $\gamma$ defined in
Eq.~(\protect\ref{eq:nucleation}),
for $\theta = 0.3$~ML
(Fig.~\protect\ref{fig:gamma})
and $\theta = 0.125$~ML.
}
\label{tab:gamma}
\end{table}

Unlike the energy barriers, 
we do not have a value for 
the attempt rate $\nu$, from atomic
scale calculations.
Instead, the {\em experimental} value
of $\nu$ is obtained by fitting the simulation and experimental 
results.
This is possible since in the simulation
$\nu$ sets a fundamental clock rate, while
the simulation results depend only on the ratio
$\nu/F$ rather then on $F$ and $\nu$ separately.
Using this property we perform the simulations for a
broad range of values of $\nu/F$ and plot the island
separation $\ell$ vs.\ $\nu/F$. 
The attempt rate $\nu$, is then obtained as the value for which
the three simulated curves in
Fig.~\ref{fig:gamma},
overlap simultaneously the three experimental curves.
It is found to be
$\nu = 1.2 \times 10^{13}$ s$^{-1}$.

In
Fig.~\ref{fig:L2invT}
we present the experimental results (\rule{2mm}{2mm})
for the island separation $\ell$ vs.\ $1/T$,
at $F = 3.21 \times 10^{-4}$~ML$\cdot$s$^{-1}$
and the corresponding simulation results ($\ast$),%
\cite{note:F}
taken at $\theta = 0.3$~ML.
The apparent good agreement between them
is a further confirmation of the correctness of $\nu$.
Both experimental and simulation curves follow the
Arrhenius behavior predicted by 
Eq.~(\ref{eq:nucleation}).
In these plots the
slope of the curve corresponds to
the activation energy 
$E_{\rm eff}$
for the rate limiting step
of the nucleation process.
The results are
$E_{\rm eff} = 0.108 \pm 0.005$
and
$0.112 \pm 0.008$~eV 
for the experimental and simulation
curves, respectively.

\begin{figure}
\vspace{-50mm}
\centerline{
\hskip -30mm
\vbox{\epsfxsize=210mm \epsfbox {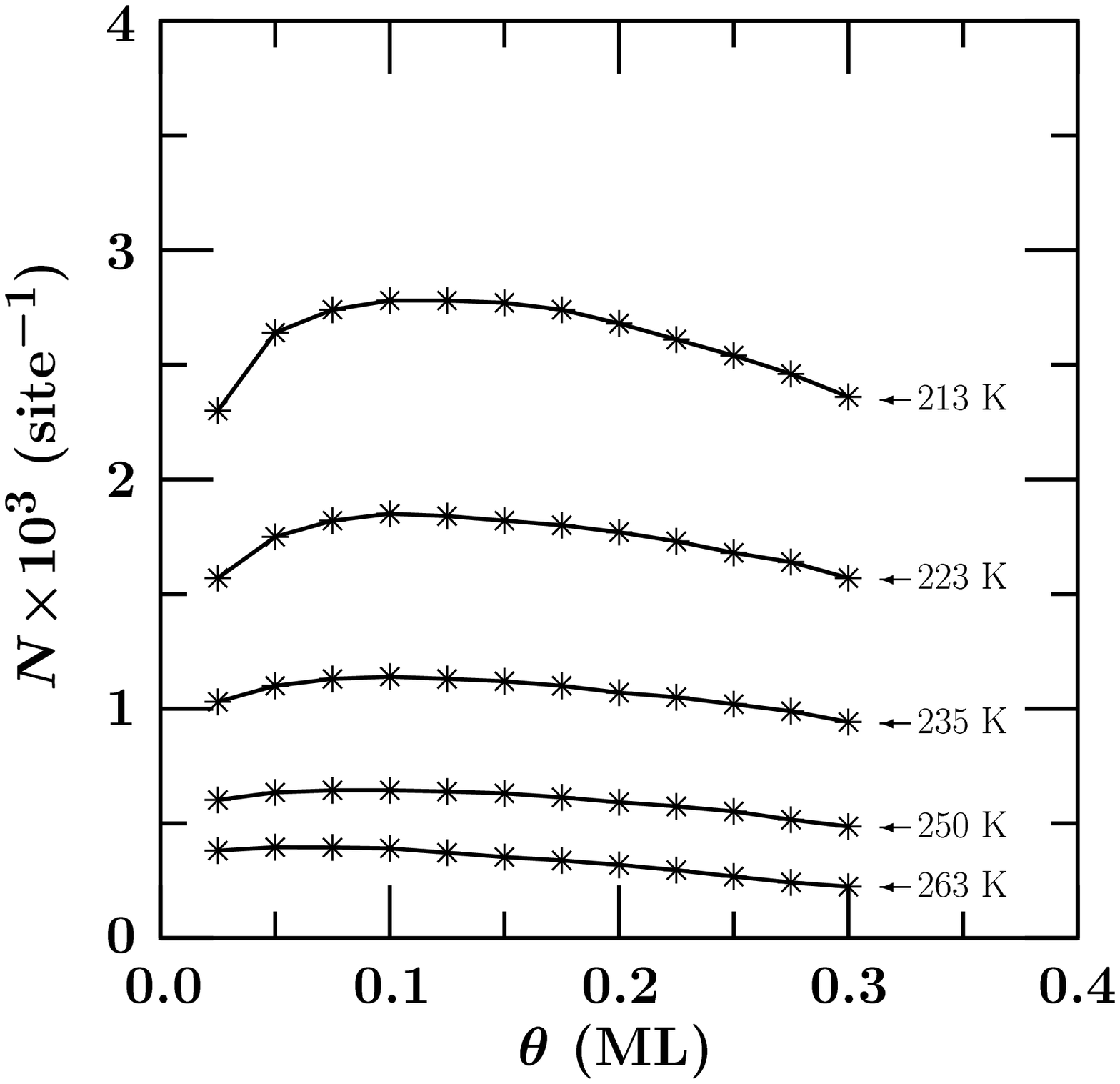}}
}
\vspace{-80mm}
\caption{
The island density $N$, vs.\ the coverage $\theta$
at different temperatures
with a flux $F = 4.8 \times 10^{-4}$~ML$\cdot$s$^{-1}$.
The solid lines are a guide to the eye.
Since all the runs begin with an empty surface,
the point $(0,0)$ is common to all the curves
although it is not drawn in the figure.
}
\label{fig:N2theta2T}
\end{figure}

In 
Fig.~\ref{fig:N2theta2T} 
we present the simulation results for
the island density $N$ vs.\ the coverage $\theta$,
for $F = 4.8 \times 10^{-4}$~ML$\cdot$s$^{-1}$
and different temperatures within the experimental range
$213$ -- $263$~K.
The data extracted from the simulations is marked by $\ast$, while
the connecting lines are only guides to the eye.
The quick rise of $N$ in the nucleation regime,
is followed by a plateau in the aggregation regime,
and then a slow decrease that is the mark of the coalescence regime.
The coalescence at the lower temperatures $T = 213$ and $223$~K,
is system-size independent and is due to island separation being
comparable to island size.
At the higher temperatures, especially $T = 263$~K,
the coalescence is mainly due to
island separation being comparable to system size,
i.e., finite size effects.
We believe that this effect reproduces to a certain extent
the effect of the steps limiting the width of the terraces
in the experimental system.%
\cite{note:terrace_width}
Our simulations indicate
that at $\theta = 0.3$~ML, where the experiments were done,
the system is already in the coalescence regime.
Therefore,
the comparison of both experiment and simulations
with theory
[Eq.~(\ref{eq:nucleation})] is misleading.

The comparison between the experimental and simulation results shows
that the set of energy barriers used here,
provides a good quantitative description
of the submonolayer growth in the low coverage regime.
We now proceed to analyze the results in the context
of available theoretical work.
In 
Table~\ref{tab:hoprates} 
we present the calculated diffusion coefficients
of adatoms as well as of small islands (dimers and trimers)
at the experimentally relevant temperatures.
The diffusion coefficient of a dimer $D_2$, is found to be comparable to
the adatom diffusion coefficient $D_1$.
The diffusion coefficients of trimers and larger islands (not shown)
are almost two orders of magnitude smaller.
Thus, only adatoms and dimers are mobile on
the surface on a time scale relevant to the growth process.
In this case the mean island separation in the {\em aggregation} regime
is given by:%
\cite{Furman1997,Biham1998S,ref:gamma4mobil}
\begin{equation}
\label{eq:nuc4dimers}
	\ell \sim \left( D_1 D_2 / F^2 \right)^{1/10}	.
\end{equation}
\begin{table}
\vskip-\lastskip
\begin{tabular}{cr@{}l@{${}\pm{}$}r@{}l
			r@{}l@{${}\pm{}$}r@{}l
				r@{}l@{${}\pm{}$}r@{}l}
	$T$ (K) &	\multicolumn{4}{c}{$D_1$ (sites s$^{-1}$)} &
			\multicolumn{4}{c}{$D_2 / D_1$} &
			\multicolumn{4}{c}{$D_3 / D_1$} 
	\\	\hline

	213 &  & 3.6 & 0.1 &  &  & 1.5 & 0.1  &  &  & 0.021  & 0.002  &
	\\
	223 &  & 11.7 & 0.4 &  &  & 1.3  & 0.1  &  &  & 0.023  & 0.002  &
	\\
	263 &  & 520.0 & 15.0 &  &  & 1.27  & 0.08  &  &  & 0.040  & 0.003  &
	\\
\end{tabular}
\caption{
The adatom diffusion coefficient
and the relative dimer and trimer diffusion coefficients,
for the three studied temperatures.
Note that through the whole temperature range the relative diffusion
coefficients almost do not change.
The diffusion coefficients of islands of size $4$ and higher
are orders of magnitude lower.
}
\label{tab:hoprates}
\end{table}
In
Ref.~\onlinecite{Biham1998S} 
it was shown that
the mobility of monomers and dimers is fully determined by 
only three energy barriers,
the single adatom hopping $E_0$,
the dimer lateral-bond breaking $E_2$
and the re-establishing of a NN bond $E_4$
(the labeling follows
Refs.~\onlinecite{Biham1998S,Mehl1999}):
\[
\unitlength 0.6mm
E_0 \! = \! E \left(
	\begin{picture}(10,6)(0,3)
		\put(4,4){\circle{4}}
		\put(4,4){\vector(1,0){4}}
	\end{picture}
	\right)
\hskip 7mm
E_2 \! = \! E \left(
	\begin{picture}(10,6)(0,3)
		\put(4,2){\circle{4}}
		\put(4,2){\vector(1,0){4}}
		\put(4,6){\circle{4}}
	\end{picture}
	\right)
\hskip 7mm
E_4 \! = \! E \left(
	\begin{picture}(10,6)(0,3)
		\put(2,2){\circle{4}}
		\put(2,2){\vector(1,0){4}}
		\put(6,6){\circle{4}}
	\end{picture}
	\right)
	.
\]
To determine also the dimer stability
one needs to specify the dimer bond-breaking energy $E_8$,
and the next-nearest neighbor bond-breaking energy $E_1$:
\begin{eqnarray*}
\unitlength 0.6mm
E_1 \! = \! E \left(
	\begin{picture}(10,6)(0,3)
		\put(2,6){\circle{4}}
		\put(6,2){\vector(1,0){4}}
		\put(6,2){\circle{4}}
	\end{picture}
	\right)
\hskip 7mm
E_8 \! = \! E \left(
	\begin{picture}(10,6)(0,3)
		\put(6,4){\circle{4}}
		\put(6,4){\vector(1,0){4}}
		\put(2,4){\circle{4}}
	\end{picture}
	\right)
	.
\end{eqnarray*}
Specifically, in the case of Cu,
$E_0 = 0.485$, $E_1 = 0.563$
(In Ref.~\onlinecite{Mehl1999} the value of this barrier was mistyped.),
$E_2 = 0.463$, $E_4 = 0.183$ and $E_8 = 0.811$~eV.
Dimer breaking can take place either from the NN configuration
with energy barrier $E_8$,
or from the next-nearest neighbor configuration
with effective energy barrier $E_2 - E_4 + E_1 = 0.843$~eV.
The high values of these barriers guarantee the dimer stability.
The adatom diffusion is a single step process,
hence, $D_1 = \nu \exp \left( - E_0 / K_{\rm B} T \right)$
[Eq.~(\ref{eq:hoppingrate})].
The dimer diffusion is a double-step process involving
a move with barrier $E_2$, followed by a move with $E_4$.
The former move is the rate-limiting move,
thus, up to a combinatorial factor of order $1$ we approximate
$D_2 = \nu \exp \left( - E_2 / K_{\rm B} T \right)$.
Since 
$E_2 \simeq E_0$ 
we expect $D_2$ and $D_1$ to be comparable,
as is indeed the case
(see Table~\ref{tab:hoprates}).
Note that this energy structure is predicted to be
common to most of the fcc(001) metal surfaces.%
\cite{Mehl1999}

Inserting $D_1$ and $D_2$ into Eq. 
(\ref{eq:nuc4dimers})
and comparing with Eq.
(\ref{eq:nucleation})
we get
\begin{equation}
\label{eq:E0+E2}
	E_{\rm eff} = (E_0 + E_2) / 10  = 0.0948 {\rm eV}	,
\end{equation}
that differs by more than $10 \%$ from the values obtained
from the simulations
($E_{\rm eff} = 0.112 \pm 0.008$~eV)
and the experiment
($E_{\rm eff} = 0.108 \pm 0.005$~eV)
presented above in
Fig.~\ref{fig:L2invT}.
We recall that the evaluation of $E_{\rm eff}$ was done at
$\theta = 0.3$~ML
where the system is already in the coalescence regime,
and therefore deviations from
Eqs.~(\ref{eq:nucleation}),(\ref{eq:nuc4dimers}) and (\ref{eq:E0+E2})
are expected.
Therefore, to
examine the scaling relations of
Eqs.~(\ref{eq:nuc4dimers})
and 
(\ref{eq:E0+E2})
we recalculate
the island separation $\ell$ vs.\ $1/T$ 
in the same temperature range
for a lower coverage of 
$\theta = 0.125$~ML
where the validity of those equations is better satisfied.
We find that 
$E_{\rm eff} = 0.096 \pm 0.003$~eV
in good agreement with
Eq.~(\ref{eq:E0+E2}).
These results are in excellent agreement with
recent calculations by Boisvert and Lewis.
\cite{Boisvert1997}

We will now revisit the calculation of $\gamma$.
According to the scaling theory
[Eq.~(\ref{eq:nuc4dimers})]
$\gamma = 2/5$ for a system with mobile monomers and dimers.%
\cite{Furman1997,ref:gamma4mobil}
The experimental results for $\gamma$
and the corresponding simulation results,
obtained at $\theta = 0.3$~ML
and presented in
Table~\ref{tab:gamma}
are significantly different from that value.
The discrepancy is removed when we recalculate $\gamma$
for $T=263$~K and a {\em much lower coverage}, $\theta = 0.125$~ML, where
Eq.~(\ref{eq:nuc4dimers})
applies.
The results of this recalculation are presented
in the rightmost column of Table~\ref{tab:gamma}.
Indeed we find $\gamma \simeq 2/5$,
in good agreement with the scaling prediction.
For $T = 213$ and $223$~K in the studied flux range
the aggregation regime is wiped out,
and the system is dominated by either nucleation or coalescence.
Therefore, deviations from the scaling prediction are expected
even for the lower coverage $\theta = 0.125$~ML,
as observed in Table~\ref{tab:gamma}.

The diffusion and coarsening of large Cu islands on Cu(001)
has recently been studied experimentally.%
\cite{Pai1997S}
The results were analyzed using MC simulation
and a model that is qualitatively consistent with our model
for the barriers relevant to large island diffusion.%
\cite{Heinonen1999S}
However, the models differ significantly in
the lateral-bond breaking barrier, $E_2$,
that is dominant in the monolayer growth through island nucleation.
Actually, the crucial factor is the difference 
$E_2 - E_0 = 0.463 - 0.485 = -0.022$~eV in our model, and
$0.52 - 0.399 = 0.121$~eV in the model of Heinonen et al.%
\cite{Heinonen1999S}
It follows that while
$D_2 / D_1 = \exp [-(E_2 - E_0) / K_{\rm B} T] $
is between 3.3 and 2.6 in the present model
for temperatures between $213$ and $263$~K,
it is between 0.001 and 0.005
in the model of Heinonen et al.%
\cite{Heinonen1999S}
Thus, according to their model dimers are {\em static}
entities on the surface giving rise to
a different scaling properties.
On the other hand,
the barriers of Shi et al.,%
\cite{Shi1996S}
$E_0 = 0.503$ and $E_2 = 0.494$~eV,
are in good agreement with ours.

In summary, we have presented a comparison between
kinetic MC simulations and experiments of Cu/Cu(001) growth
in the submonolayer regime.
The available experimental data is for $\theta = 0.3$~ML
and $T = 213$, $223$ and $263$~K.
At this coverage a good agreement between simulation and experimental
results is found.
However, these results deviate from mean-field predictions,
a fact that we attribute to the system being away from the aggregation regime
at $\theta = 0.3$~ML.
Indeed, repeating the analysis of the simulation results
for $\theta = 0.125$~ML we find
$\gamma \sim 0.4$
and
$E_{\rm eff} = 0.096 \simeq (E_0 + E_2)/10$~eV
consistent with mean-field predictions when
only adatoms and dimers are mobile on the surface.
We conclude that for Cu/Cu(001), in the studied temperature range,
only adatoms and dimers are mobile.
Furthermore, mean-field predictions apply at $\theta \sim 0.1$~ML
but not at much higher coverage.
Finally, the comparison between simulation and experiment,
augmented with a scaling procedure,
enables us to depict the attempt rate
$\nu = 1.2 \cdot 10^{13}$ s$^{-1}$
of the experimental system.

We thank Hanoch Mehl for helpful discussions.
Experimental portions of this work were supported by Oak
Ridge National Laboratory, which is managed by Lockheed Martin Energy Research
Corporation for the U.S. Department of Energy under Contract No.
DE-AC05-96OR22464.

\onecolumn

\bibliography{general,surface,paper}

\begin{thebibliography}{10}

\bibitem{ref:STM4T}
{ Y. W. Mo {\it et al.}, Phys. Rev. Lett. {\bf 66}, 1998 (1991); J. A.
  Stroscio, D. T. Pierce and R. A. Dragoset, Phys. Rev. Lett. {\bf 70}, 3615
  (1993); H. Brune {\it et al.}, Phys. Rev. Lett. {\bf 73}, 1955 (1994); F.
  Tsui {\it et al.}, Phys. Rev. Lett. {\bf 76}, 3164 (1996) }.

\bibitem{ref:STM4FT}
{ C.-M. Zhang {\it et al.}, Surf. Sci. {\bf 406}, 178 (1998); S. G\"{u}nther
  {\it et al.}, Phys. Rev. Lett. {\bf 73}, 553 (1994); T. R. Linderoth {\it et
  al.}, Phys. Rev. Lett. {\bf 77}, 87 (1996) }.

\bibitem{Stroscio1994}
{J. A. Stroscio and D. T. Pierce}, Phys. Rev. B {\bf 49},  8522  (1994).

\bibitem{Zuo1994S}
{J.-K. Zuo {\it et al.}}, Phys. Rev. Lett. {\bf 72},  3064  (1994).

\bibitem{Durr1995}
{H. D\"{u}rr, J. F. Wendelken and J.-K. Zuo}, Surf. Sci. {\bf 328},  {L527}
  (1995).

\bibitem{Swan1997S}
{A. K. Swan {\it et al.}}, Surf. Sci. {\bf 391},  {L1205}  (1997).

\bibitem{Bardotti1998S}
{L. Bardotti {\it et al.}}, Phys. Rev. B {\bf 57},  12544  (1998).

\bibitem{Ernst1992}
{H. J. Ernst, F. Fabre and J. Lapujoulade}, Phys. Rev. B {\bf 46},  1929
  (1992).

\bibitem{ref:nucleation}
{ J. A. Venables, {Phil. Mag.} {\bf 27}, 697 (1973); S. Stoyanov and D.
  Kashchiev, {Current Topics in Material Science} {\bf 7}, 70 (1981) }.

\bibitem{note:terrace_width}
{ This is comparable to the average terrace width in the experiment, that was
  estimated to be 70 nm \cite{Zuo1994S,Durr1995,Swan1997S} }.

\bibitem{Voter1986}
{A.F. Voter}, Phys. Rev. B {\bf 34},  6819  (1986).

\bibitem{Furman1997}
{I. Furman and O. Biham}, Phys. Rev. B {\bf 55},  7917  (1997).

\bibitem{Biham1998S}
{O. Biham {\it et al.}}, Surf. Sci. {\bf 400},  29  (1998).

\bibitem{note:barriers}
{The $3 \times 3$ environment consists of $7$ sites adjacent to the initial and
  final sites of the hopping move. This gives rise to $2^7 = 128$ local
  configurations }.

\bibitem{Daw1983}
{M. S. Daw and M. I. Baskes}, Phys. Rev. Lett. {\bf 50},  1285  (1983).

\bibitem{Mehl1999}
{H. Mehl, O. Biham, I. Furman and M. Karimi}, Phys. Rev. B {\bf 60},  2106
  (1999).

\bibitem{note:coverage}
{ This value is above the minimum measurable coverage of $\theta = 0.2$~ML. For
  the data obtained at higher coverage $\theta > 0.6$~ML \cite{Zuo1994S} the
  suppression of second-layer nucleation in our simulations may not be
  appropriate. Also, scaling theories do not apply in this regime. Therefore,
  we concentrated only on the $\theta = 0.3$ data }.

\bibitem{note:Sshape}
{ Note that the experimental curve at $T=223$K has a slight S-shaped modulation
  (see also Fig.\ 3 in Ref.\ \onlinecite{Zuo1994}). This gives rise to the
  deviation in the exponent $\gamma$ compared to the simulation results.
  However, the actual differences in the data points between the experiment and
  simulation are rather small }.

\bibitem{note:range4fits}
{ In the present analysis we have used the broadest possible range common to
  all data sets, that provides simultaneous high-quality fits for all of them.
  In Refs.~\onlinecite{Zuo1994S,Swan1997S}, however, a narrower flux range was
  employed, giving rise to different values }.

\bibitem{ref:crossover}
{ M. C. Bartelt {\it et al.}, Phys. Rev. B {\bf 53}, 4099 (1996); M.N. Popescu,
  J.G. Amar and F. Family, Phys. Rev. B {\bf 58}, 1613 (1998) }.

\bibitem{note:F}
{For technical reasons, we used in the simulation $F = 4.8 \times 10^{-4}$
  instead of $3.21 \times 10^{-4}$~ML$\cdot$s$^{-1}$. According to
  Eq.~(\ref{eq:nucleation}), this would multiply the island separation by a
  factor of $ (4.8 / 3.21)^{\gamma /2} $. From Table~\ref{tab:gamma} we see
  that this amounts to a deviation of $5 \%$ -- $15 \%$, which is approximately
  the deviation that is apparent from Fig.~\ref{fig:L2invT}}.

\bibitem{ref:gamma4mobil}
{ J. Villain {\it et al.}, J. Phys. I (France) {\bf 2}, 2107 (1992); {D. E.
  Wolf}, in {\em {Scale Invariance, Interfaces and Non-Equilibrium Dynamics}},
  Vol.~344 of {\em NATO Advanced Study Institute, Series B: Physics}, edited by
  {M. Droz, A. J. McKane, J. Vannimenus and D. E. Wolf} (Plenum, New York,
  1994) }.

\bibitem{Boisvert1997}
{G. Boisvert and L. J. Lewis}, Phys. Rev. B {\bf 56},  7643  (1997).

\bibitem{Pai1997S}
{W.W. Pai {\it et al.}}, Phys. Rev. Lett. {\bf 79},  3210  (1997).

\bibitem{Heinonen1999S}
{J. Heinonen {\it et al.}}, Phys. Rev. Lett. {\bf 82},  2733  (1999).

\bibitem{Shi1996S}
{Z.-P. Shi {\it et al.}}, Phys. Rev. Lett. {\bf 76},  4927  (1996).

\bibitem{Zuo1994}
{J.-K. Zuo, J. F. Wendelken, H. D\"{u}rr, and C.-L. Liu}, Phys. Rev. Lett. {\bf
  72},  3064  (1994).

\end{thebibliography}
\bibliographystyle{prsty}


\pagestyle{empty}
\noindent

\noindent

\noindent

\end{document}